# Shift, Scale and Rotation Invariant Multiple Object Detection using Balanced Joint Transform Correlator

XI SHEN,[1] JULIAN GAMBOA,[1] TABASSOM HAMIDFAR,[1] SHAMIMA A. MITU,[1] AND SELIM M. SHAHRIAR[1,2]

[1]*Department of Electrical and Computer Engineering, Northwestern University, Evanston, IL 60208, USA*
[2]*Department of Physics and Astronomy, Northwestern University, Evanston, IL 60208, USA*
* *shahriar@northwestern.edu*

**Abstract:** The Polar Mellin Transform (PMT) is a well-known technique that converts images into shift, scale and rotation invariant signatures for object detection using opto-electronic correlators. However, this technique cannot be properly applied when there are multiple targets in a single input. Here, we propose a Segmented PMT (SPMT) that extends this methodology for cases where multiple objects are present within the same frame. Simulations show that this SPMT can be integrated into an opto-electronic joint transform correlator to create a correlation system capable of detecting multiple objects simultaneously, presenting robust detection capabilities across various transformation conditions, with remarkable discrimination between matching and non-matching targets.

## 1. Introduction

Real-time target recognition is a fundamental part of modern computational systems, attracting significant attention due to its widespread applications. While recent advances in machine learning algorithms such as YOLOv10 [1] have enhanced detection capabilities, optical signal processing systems offer inherent advantages through analog mechanisms that perform computations at the speed of light propagation [2–6], regardless of image resolution or complexity.

All-optical correlators employ converging lenses to generate the Fourier transforms (FTs) of two input images, using either nonlinear materials or holographic filters to produce their conjugate product, which in turn produces the cross-correlation after passing through another converging lens. While these systems can operate at the speed of light, their nonlinear materials or holographic filters make them largely incompatible with systems that require real-time processing. In contrast, hybrid opto-electronic correlators [7–9] use Focal Plane Arrays (FPAs) as the nonlinear medium that captures the FTs of the input images, subsequently employing Spatial Light Modulators (SLMs) to project their conjugate product back into the optical domain. The exact layout and function of the FPAs and SLMs depends on the exact hybrid opto-electronic architecture that is employed. In an opto-electronic Joint Transform Correlator (JTC), for example, a single SLM simultaneously projects the two input images side by side, which together get FT'd using a lens. The intensity of the interference between these two FTs, labelled the Joint Power Spectrum (JPS), is captured by an FPA, subsequently projected on an output SLM, and finally passed through another lens to produce the cross-correlation. Recently, a Balanced Opto-electronic Joint Transform Correlator (BOJTC) architecture was shown to improve on this design by [10] introducing additional electronic processing to remove the unnecessary self-intensity terms from the JPS, thereby expanding the detection quality [11] with higher discrimination capability.

Regardless of which opto-electronic correlation architecture is used, a significant limitation of these correlation systems is their inability to detect targets that have undergone scaling or rotation relative to the reference image. To address this constraint, the Polar Mellin Transform (PMT) [12,13] is an excellent pre-processing technique. The PMT can be computed as the log-polar transform (LPT) of the magnitude of the FT, thereby leveraging Fourier optics to enhance processing speed while retaining most of the relevant information in a manner that enables correlation despite shift, scale, and rotation differences. An experimental opto-electronic PMT pre-processing (OPP) [14,15] stage was recently constructed that operates at the same speed as the remainder of the opto-electronic correlation system, producing an overall operational speed of 720 frames per second. While the PMT demonstrates exceptional performance for single object detection, multiple object scenarios present significant challenges due to interference patterns between FTs of different objects. These challenges significantly degrade detection reliability, which this paper addresses by systematically dividing the input into segments and processing each independently.

When multiple objects are present within the same frame, the overall frame dimensions and resolution requirements necessarily increase to accommodate all targets with sufficient detail, resulting in substantially greater computational demands. This expanded data volume introduces significant challenges for digital processing systems due to the quadratic relationship between frame size and computational complexity. Conventional approaches could mitigate this challenge by sequentially processing individual objects or regions of interest; however, this sequential methodology inevitably extends the total processing time proportionally to the number of objects present. Opto-electronic correlators offer a compelling solution to this challenge by enabling parallel computation for multiple targets with high operational speed. All target objects can be displayed within a single input frame and processed simultaneously in one operational cycle, constrained primarily by the spatial resolution of the optical components rather than by computational limitations. This makes the segmented PMT a particularly good fit with architectures such as the BOJTC.

The rest of this paper is structured as follows. Section 2 introduces the mathematical foundations of the Segmented Polar Mellin Transform (SPMT). In Section 3, we present a comprehensive analysis of the simulation results for multiple object detection using SPMT. Section 4 shows the integration of SPMT with the BOJTC, providing detailed simulation results of this investigation. Finally, Section 5 concludes the findings of this paper, and discusses the limitations and future work.

## 2. Segmented Polar Mellin Transform

### 2.1 Polar Mellin Transform

The mathematical foundation of the PMT begins with the calculation of the LPT of the magnitude of the FT'd image. In principle, this is a coordinate mapping process where the image is transformed from a $(x, y)$ coordinate frame to a $(\rho, \theta)$ coordinate frame projected onto a cartesian plane. After choosing coordinates such that the center of the FPA is the zero point ($x = 0, y = 0$), the calculation of the LPT starts as follows. First, the horizontal axis, $\rho$, is defined as the logarithm of the radial coordinate, $r$, of each pixel in the FT plane: $\rho = \ln\left(\frac{r}{r_0}\right) = \ln\left(\frac{1}{r_0}\sqrt{x^2 + y^2}\right)$, where $r_0$ is a the minimum radius to be transformed, thus defining the lowest frequency of interest in the system. The mapping process is restricted to the domain of $r \geq r_0$, thus ensuring that the minimum value of $\rho$ is zero. In effect, this restriction of the domain represents blocking the DC component at the center of the FT'd image. Next, the vertical axis, $\theta$, is defined as the angular coordinate corresponding to each pixel in the FT: $\theta = \text{atan}\left(\frac{y}{x}\right)$, where $\theta$ will range from 0 to $2\pi$ radians.

The PMT offers inherent shift invariance, because the shift information is encoded within the phase of its FT, and eliminated upon detection of the intensity of the FT. This, however, also means that the original position information of the target cannot be recovered from its PMT. For rotated objects, the corresponding FT rotates by the same amount, $\varphi$, resulting in a linear shift along the vertical $\theta$-axis in the PMT. In instances where the object is scaled by a factor of $\alpha$, the corresponding FT will be scaled by a factor of $1/\alpha$. Additionally, because $\rho(r/\alpha) = \ln\left(\frac{r}{r0}\right) - \ln(\alpha) = \rho(r) - \ln(\alpha)$, this would result in a linear shift along the $\rho$-axis by an amount $\ln(\alpha)$. Thus, the value of $\alpha$ can be inferred from this shift. These properties allow the system to detect objects regardless of their position, orientation, or size.

## 2.2 Segmented Polar Mellin Transform

While the PMT is useful for single object detection, multiple object scenarios present significant challenges. When multiple objects are present in a frame, their FTs overlay, creating interference patterns that complicate individual object detection. As illustrated in Figure 1, a single square object shown in Figure. 1.A(1) produces a FT (Figure.

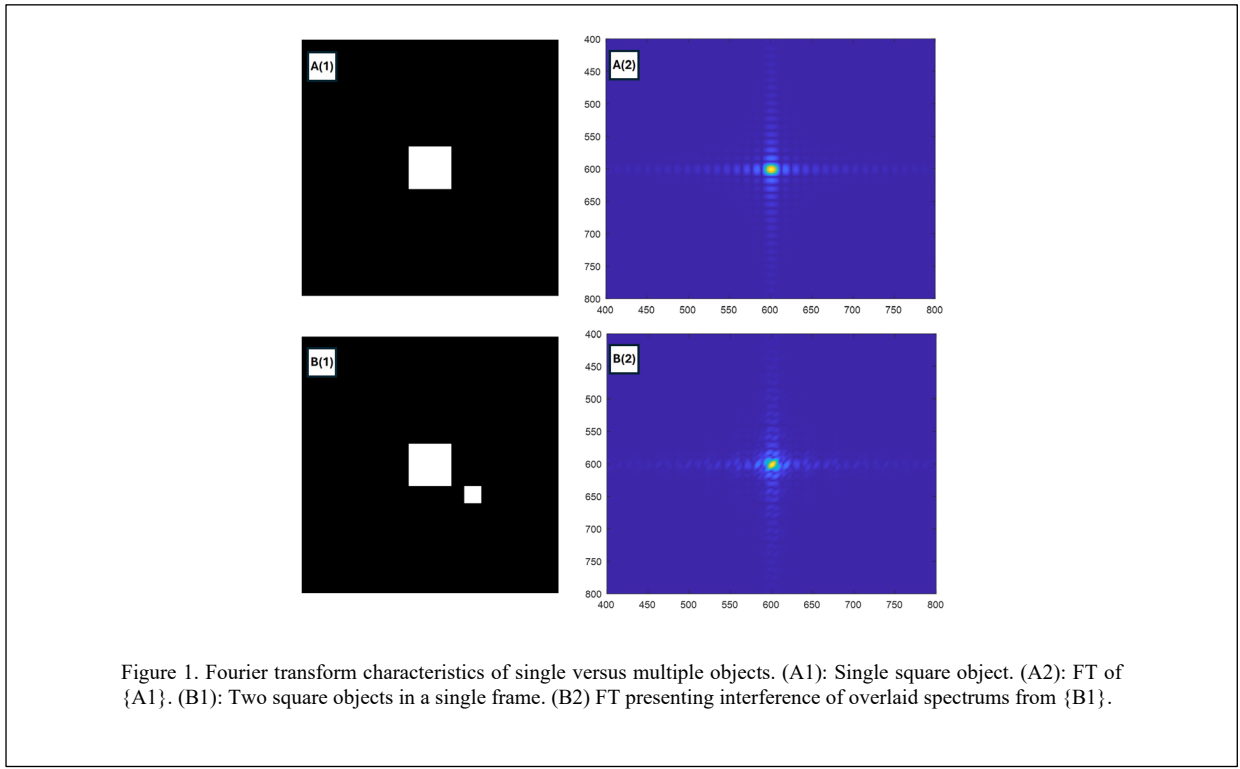

Figure 1. Fourier transform characteristics of single versus multiple objects. (A1): Single square object. (A2): FT of {A1}. (B1): Two square objects in a single frame. (B2) FT presenting interference of overlaid spectrums from {B1}.

1.A(2)) that appears as a well-defined two-dimensional sinc function. When two square objects are present in the same frame as shown in Figure. 1.B(1), they produce interference patterns (Figure. 1.B(2)) that preclude the PMT from being carried out for each object.

This limitation also affects spatio-temporal correlators, wherein the PMT and a one-dimensional equivalent have recently been proposed as a means to perform speed-invariant video detection within large databases. In this case, when the temporal signal of each pixel undergoes the FT, event information from multiple signals become entangled, making independent extraction difficult or impossible [16].

To overcome these challenges, we propose an SPMT approach as a preprocessing technique that effectively handles multiple object detection in controlled scenarios such as database searches. This method systematically divides the input frame into multiple sections or regions of interest, establishing the center of each section as a local zero point for coordinate transformation. By conducting the PMT independently for each section, the system isolates individual objects and transforms them into Shift, Scale, and Rotation Invariant (SSRI) signatures. After this preprocessing stage, the transformed image sections can be fed into an opto-electronic correlator for the actual detection process. The BOJTC is particularly well-suited for this application as it enables simultaneous correlation between the preprocessed input query image and multiple reference objects with higher precision than the opto-electronic JTC.

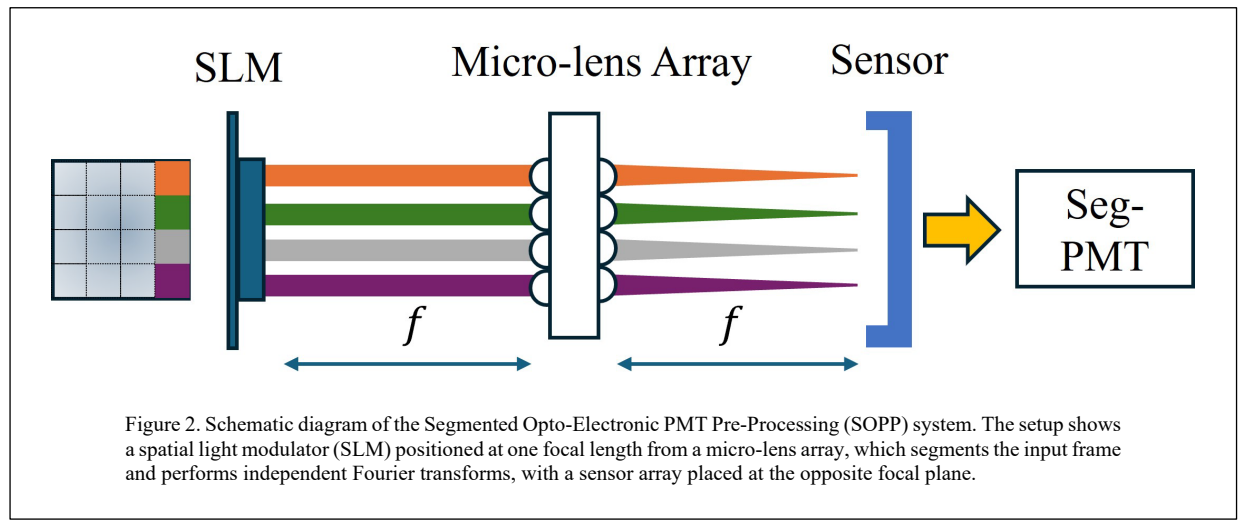

Figure 2. Schematic diagram of the Segmented Opto-Electronic PMT Pre-Processing (SOPP) system. The setup shows a spatial light modulator (SLM) positioned at one focal length from a micro-lens array, which segments the input frame and performs independent Fourier transforms, with a sensor array placed at the opposite focal plane.

The complete detection pipeline thus consists of two main stages: first, the Segmented Opto-electronic PMT Pre-processing (SOPP) step converts the input image containing multiple objects into a format where each object is represented in a SSRI manner within its own segment; second, the BOJTC performs the correlation between these preprocessed segments and the reference object database. When a match is found, the correlator produces distinct correlation peaks that correspond to the detected objects. The locations of these correlation peaks provide information about the objects' presence and their transformations relative to the reference, while avoiding interference issues that would occur with a conventional, non-segmented approach.

Figure 2 illustrates our proposed SOPP system architecture. This design leverages the parallel processing capabilities of Fourier optics by implementing a micro-lens array to simultaneously segment the input frame and perform FTs on each section independently. In the optical setup, an SLM displaying the input image is positioned at one focal length from the micro-lens array. The array consists of multiple lenses, each aligned to a specific region of this SLM. On the opposite side, an FPA is placed at the focal plane to capture the resulting FTs. Following the optical FT stage, the LPT is applied to the magnitude of each segment's FT, completing the PMT operation for each image section independently.

### 3. Simulation Analysis of the Segmented Polar Mellin Transform

*3.1 Simulation framework*

To evaluate the proposed SPMT approach for multi-object SSRI detection, a series of images were prepared such that they contained a variety of scaling and rotation factors, with both matched and mismatched images.

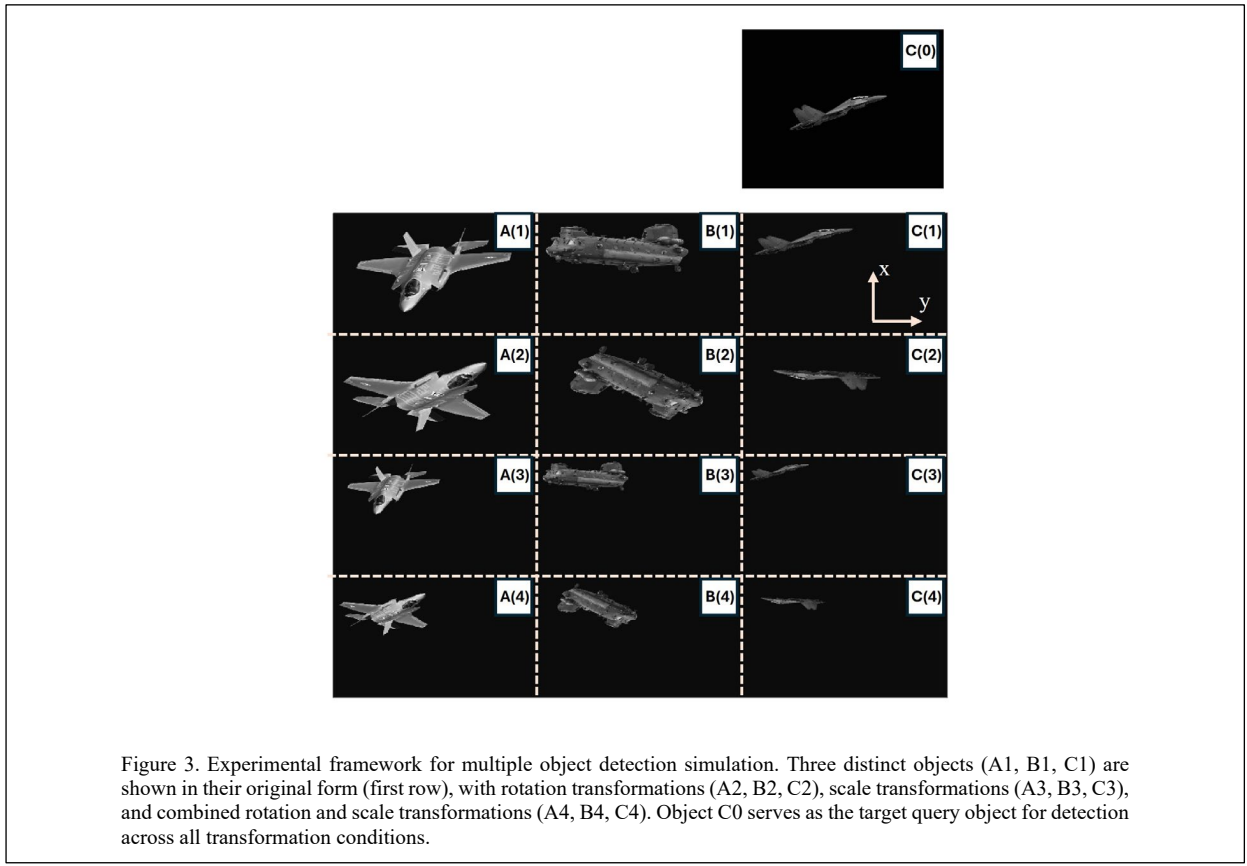

Figure 3. Experimental framework for multiple object detection simulation. Three distinct objects (A1, B1, C1) are shown in their original form (first row), with rotation transformations (A2, B2, C2), scale transformations (A3, B3, C3), and combined rotation and scale transformations (A4, B4, C4). Object C0 serves as the target query object for detection across all transformation conditions.

Figure 3 presents the experimental framework designed for our multiple object detection simulation. The test frame contains three distinct object types arranged in a grid format to systematically evaluate the system's performance under various transformation conditions.

The first row of Figure 3 displays the reference objects in their original form: Figure 3.(A1), (B1), and (C1) represent three different object types with distinct visual characteristics. The subsequent rows demonstrate these objects under different transformation conditions. The second row – Figure 3.(A2), (B2), and (C2) – shows the objects after rotation transformations while maintaining their original scale. The third row – Figure 3.(A3), (B3), and (C3) – presents the objects with scale variations while preserving their original orientation. The fourth row – Figure 3.(A4), (B4), and (C4) – combines both scaling and rotation transformations to test the system's invariance to multiple transformations.

For this simulation, the object depicted in Figure 3.(C0) serves as our target query object. The system is expected to identify matches exclusively along column C, regardless of the transformations applied to the target object. This design enables us to evaluate both the system's detection accuracy and its ability to reject non-matching objects. To ensure experimental consistency and eliminate intensity-based biases, all objects within each section are normalized to have equivalent power levels.

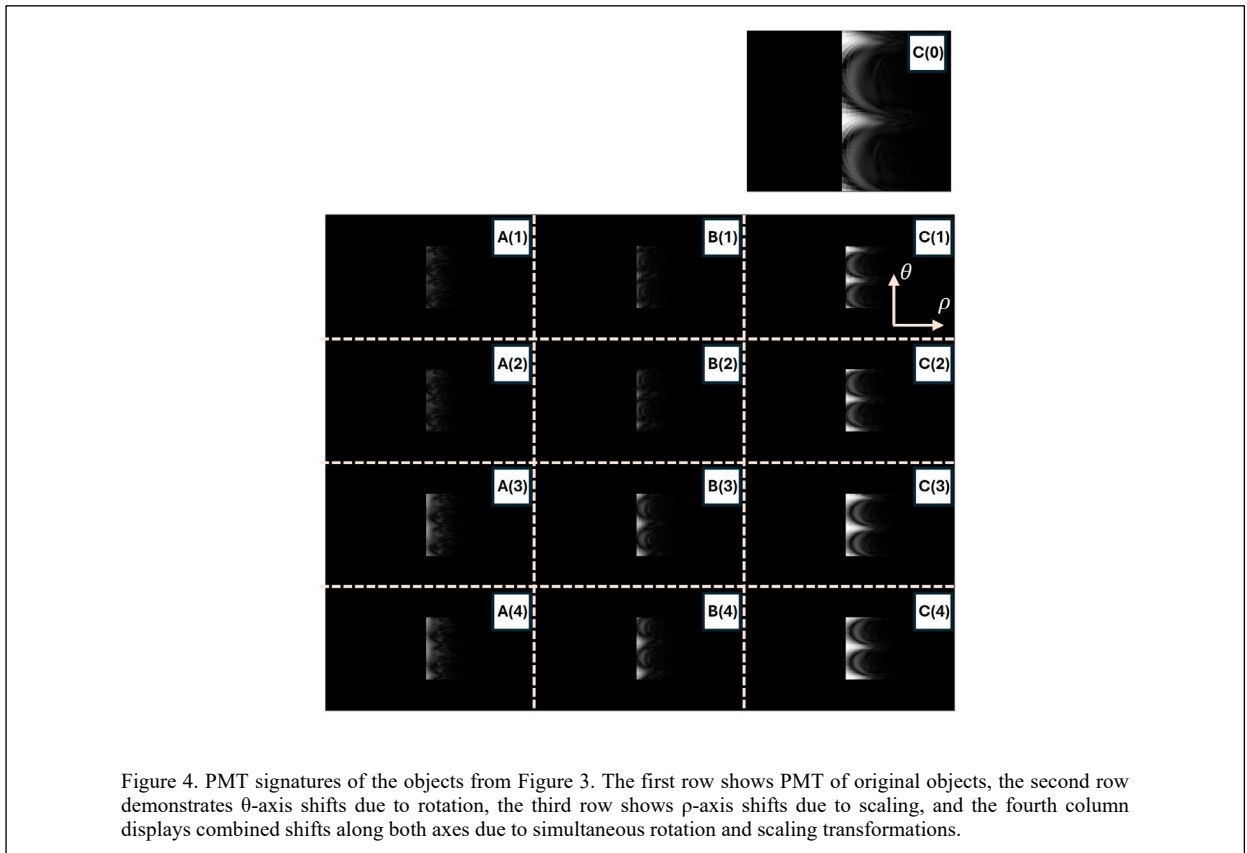

Figure 4. PMT signatures of the objects from Figure 3. The first row shows PMT of original objects, the second row demonstrates θ-axis shifts due to rotation, the third row shows ρ-axis shifts due to scaling, and the fourth column displays combined shifts along both axes due to simultaneous rotation and scaling transformations.

To detect objects with shifts, scaling, and rotations across the segmented framework, we apply the PMT individually to each section depicted in Figure 3. The resulting PMT representations are displayed in Figure 4, which illustrates the transformations using a new coordinate system with $\rho$ and $\theta$ axis.

The first row of Figure 4 shows the PMT of the original reference images. The second row demonstrates how rotation transformations display as linear shifts along the θ axis within the PMT domain. This vertical displacement corresponds directly to the angle of rotation applied to the original object. In the third row, we observe the effect of scaling transformations on the PMT representation. When the original images are reduced in size, as implemented in our experimental design, the PMT exhibits a linear shift along the $\rho$ axis. Specifically, the rightward (positive) shift in the PMT directly correlates with the reduction in scale of the original object. The fourth row presents the most complex case, where both scaling and rotation are applied to the original objects. Here, the PMT displays coordinate shifts along both the $\rho$ and $\theta$ axes, demonstrating the transform's ability to decompose complex transformations into separable linear displacements within its domain.

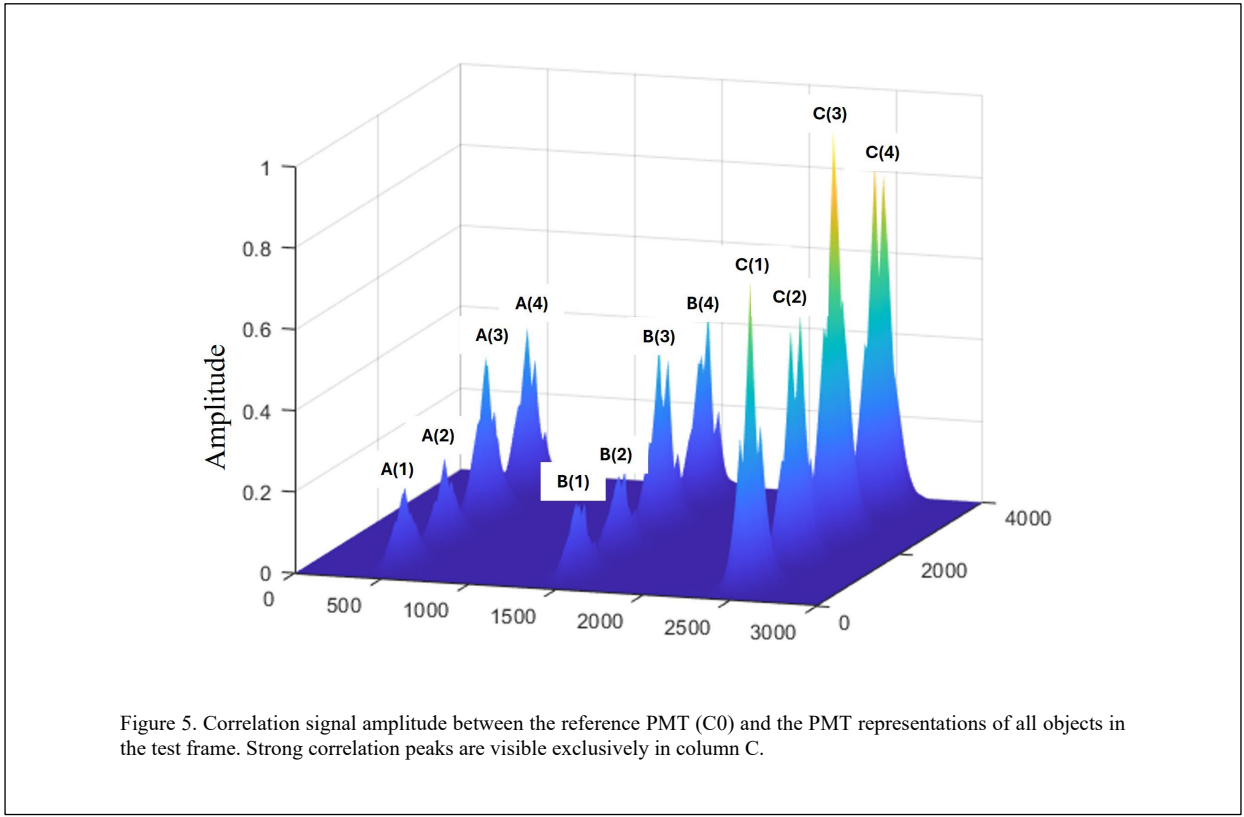

Figure 5. Correlation signal amplitude between the reference PMT (C0) and the PMT representations of all objects in the test frame. Strong correlation peaks are visible exclusively in column C.

For detection purposes, we correlate each PMT representation from the segmented frame with the PMT of our target query object shown in C(0). This correlation process identifies matching objects regardless of their shift, scale, or rotation, producing distinct correlation peaks for positive matches while maintaining low correlation values for non-matching objects.

*3.2 Correlation Analysis*

Figure 5 presents the correlation signals generated by the query PMT with the PMT of each segment in the test frame. The results demonstrate a consistently strong correlation exclusively within column C, which contains our target object. This pattern of elevated correlation values confirms the system's ability to accurately identify the target object despite various transformations applied to it.

The cross-correlation values for objects in columns A and B remain significantly lower, indicating the system's robust discrimination capability between non-matching objects. This selectivity is crucial for practical applications where false positives must be minimized while maintaining high detection sensitivity.

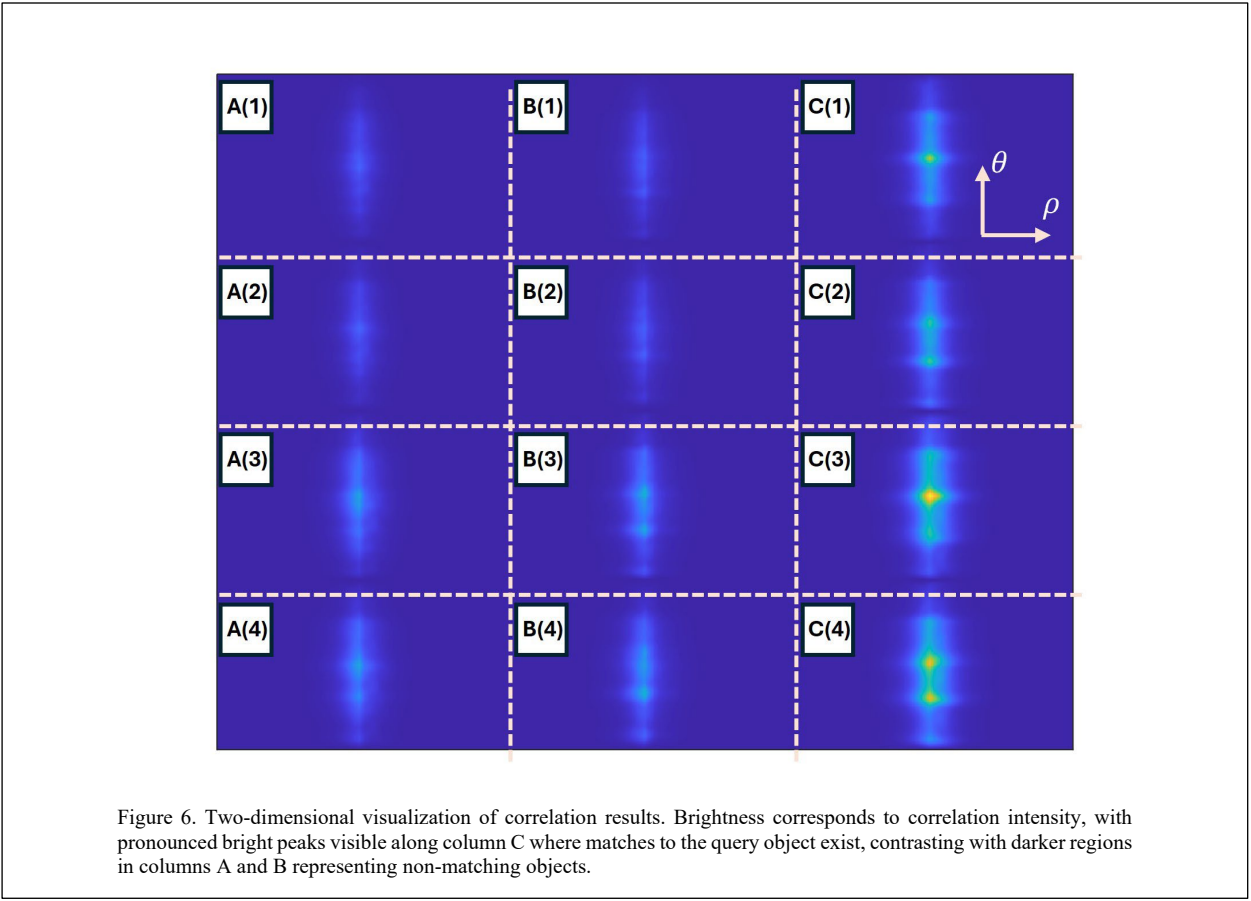

Figure 6. Two-dimensional visualization of correlation results. Brightness corresponds to correlation intensity, with pronounced bright peaks visible along column C where matches to the query object exist, contrasting with darker regions in columns A and B representing non-matching objects.

Figure 6 provides a two-dimensional visualization of the correlation results. The correlation intensity is represented as a brightness map, where higher correlation values correspond to brighter regions. This representation clearly illustrates the pronounced correlation peaks along column C, where matches the query object exist. The bright correlation peaks in column C and the relatively dark regions corresponding to columns A and B further validates the system's discrimination performance.

The consistency of detection across different transformation conditions—original, rotated, scaled, and combined transformations—demonstrates the effectiveness of the SPMT approach for SSRI multiple object detection.

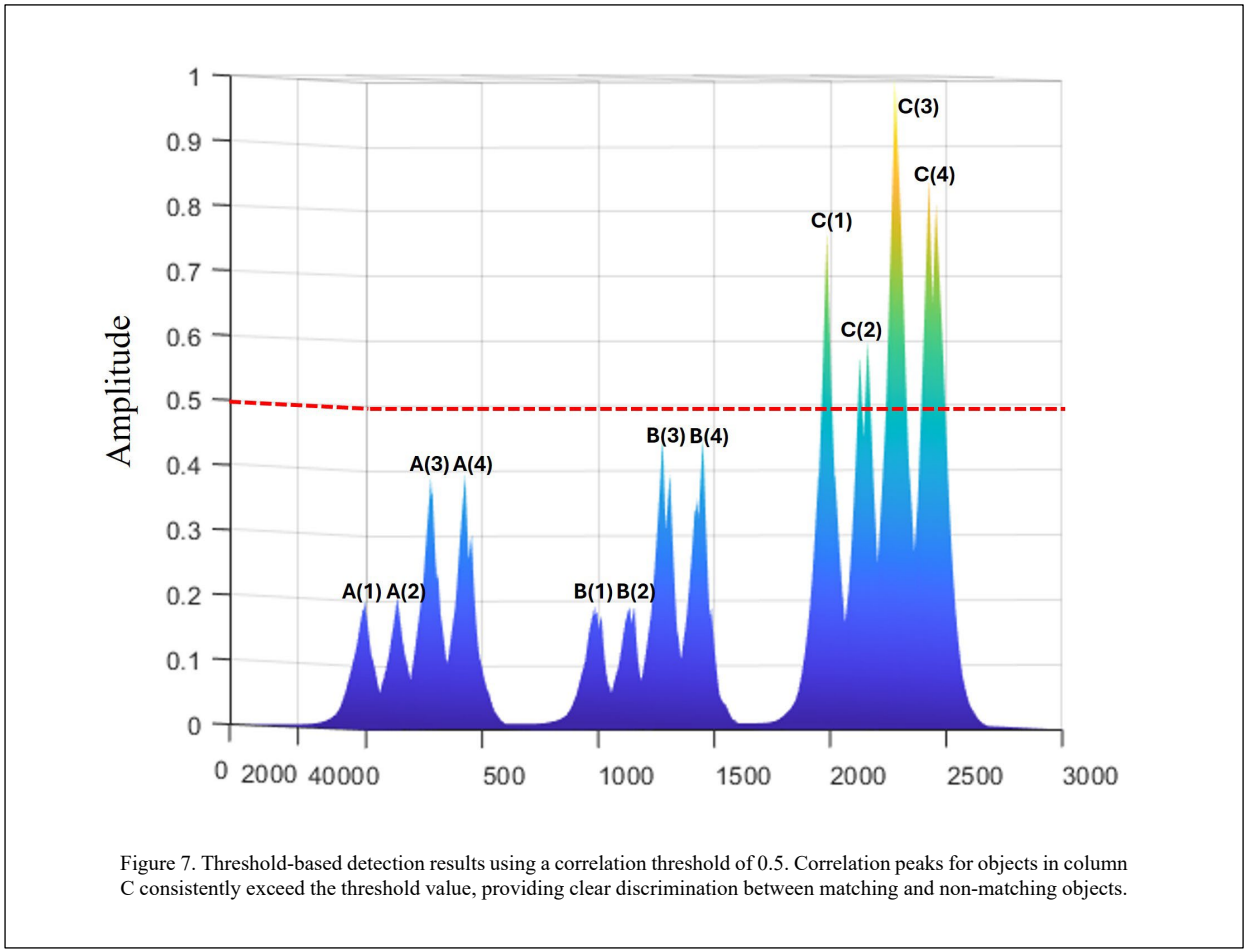

Figure 7. Threshold-based detection results using a correlation threshold of 0.5. Correlation peaks for objects in column C consistently exceed the threshold value, providing clear discrimination between matching and non-matching objects.

To establish a quantitative detection criterion, we implemented a correlation threshold value of 0.5 for positive object identification. Figure 7 illustrates the effectiveness of this threshold in discriminating between matching and non-matching objects. The correlation peaks for objects in column C consistently exceed the 0.5 threshold value, clearly identifying them as positive matches to our query object.

After applying the threshold filter to the correlation plane, we can examine the distinctive characteristics of the correlation peaks for different transformation conditions, as illustrated in Figure 8. In Figure 8.C(1), the autocorrelation of the reference object with itself produces a well-defined, centrally located peak with high amplitude. This symmetrical, singular peak represents the baseline correlation pattern for an exact match without any transformations.

When the system detects a rotated version of the target object, as shown in Figure 8.C(2), the correlation peak exhibits a characteristic pattern. This split into two distinct peaks occurs due to the circular nature of the angular coordinate in the PMT domain, where it wraps around the full $2\pi$ range. Specifically, one peak exhibits a vertical shift of $\varphi$ (the rotation angle) while the other displays a vertical shift of $2\pi-\varphi$, creating a complementary pair of peaks.

For scaled objects, Figure 8.C(3) demonstrates a horizontal displacement of the correlation peak. The rightward shift of the peak position along the horizontal axis corresponds to the reduction in scale of the detected object relative to the reference. This displacement aligns well with our expectations.

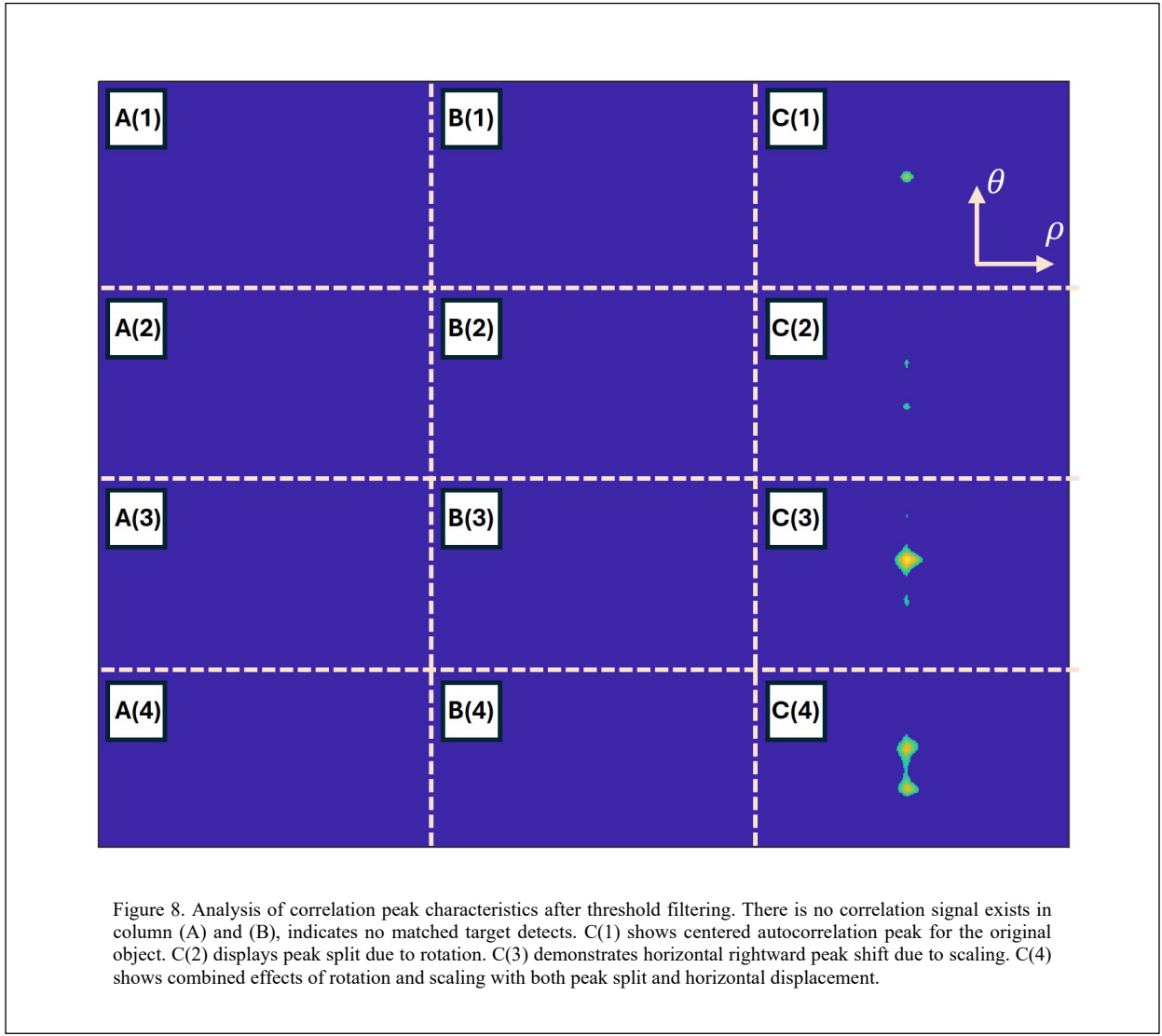

Figure 8. Analysis of correlation peak characteristics after threshold filtering. There is no correlation signal exists in column (A) and (B), indicates no matched target detects. C(1) shows centered autocorrelation peak for the original object. C(2) displays peak split due to rotation. C(3) demonstrates horizontal rightward peak shift due to scaling. C(4) shows combined effects of rotation and scaling with both peak split and horizontal displacement.

In the most complex case, Figure 8.C(4) displays the correlation pattern for an object that has undergone both scaling and rotation. Here, we observe a combination of the previous effects: the peak has both split into two separate peaks due to rotation and shifted rightward due to scaling. This distinctive pattern confirms the system's ability to detect multiple simultaneous transformations through the analysis of correlation peak characteristics.

## 4. SSRI Multiple targets recognition using BOJTC

This section demonstrates the application of the BOJTC [10] for multiple object detection with invariance to shift, scale, and rotation. The BOJTC provides an efficient method for parallel processing of multiple correlation operations, enabling simultaneous detection of multiple objects. Figure 9 illustrates the display input for the BOJTC system. The display frame is organized into two primary sections: the left section contains the original query PMT image derived from our target object, while the right section accommodates the multiple object PMT images obtained from the segmented PMT input frame. This spatial arrangement produces the joint FT of both the reference and query images simultaneously, which would produce the JPS when detected on an FPA.

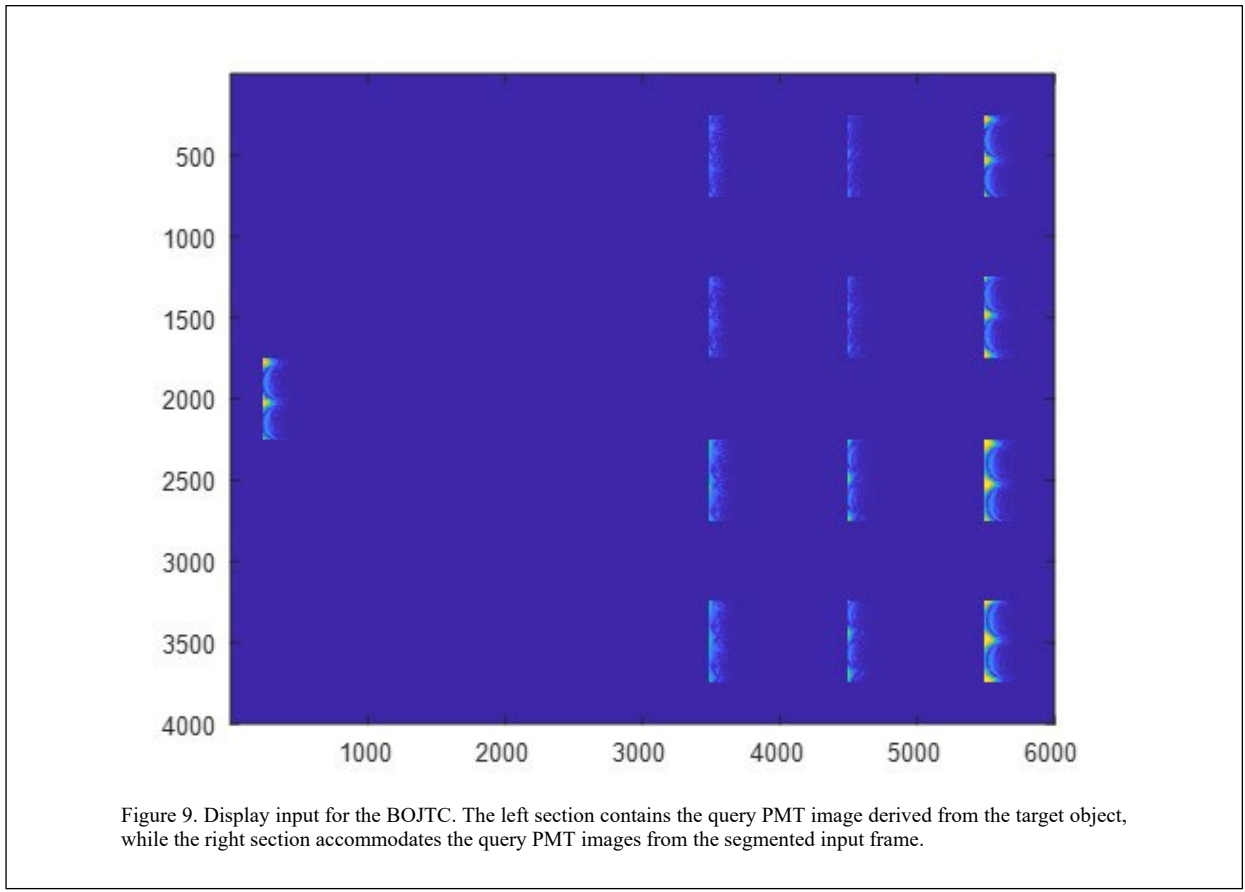

Figure 9. Display input for the BOJTC. The left section contains the query PMT image derived from the target object, while the right section accommodates the query PMT images from the segmented input frame.

Figure 10 shows the JPS of the example shown in Figure 9, which contains several components: the self-intensity terms from the reference and query images individually, and their conjugate-product. In a conventional JTC approach, the self-intensity terms can introduce unwanted power offsets that reduce the resolution, and thus the discrimination capability, of the system. To enhance detection precision, we implement the BOJTC technique, which involves subtracting the self-intensity of the reference and query images from the JPS. The resulting balanced power spectrum contains only the information that yields the cross-correlation, which facilitates high-resolution detection with improved accuracy for target recognition applications.

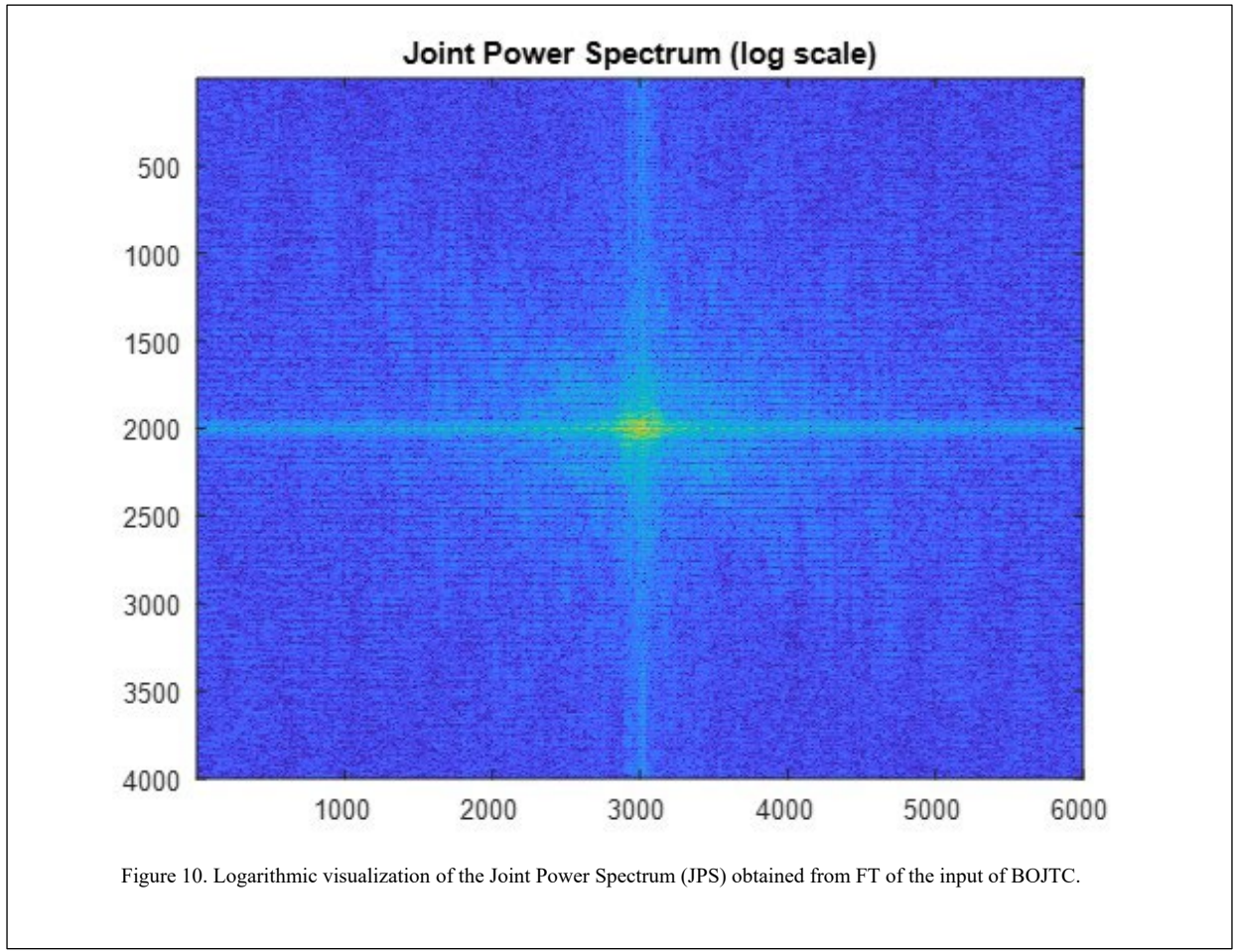

Figure 10. Logarithmic visualization of the Joint Power Spectrum (JPS) obtained from FT of the input of BOJTC.

Applying the inverse FT to the balanced JPS produces the correlation output shown in Figure 11. This output exhibits well-defined correlation patterns that demonstrate the system's detection capabilities. The correlation pattern displayed on the left side of Figure 11 shows exceptional alignment with the mathematically computed correlation results presented in our earlier analysis.

The distinct correlation peaks visible in the output plane correspond to the locations where the query images match the reference image, taking into account any shifts, rotations, or scaling transformations. These results demonstrate that the integration of the SPMT with the BOJTC creates an effective framework for multiple object detection with invariance to shift, scale, and rotation.

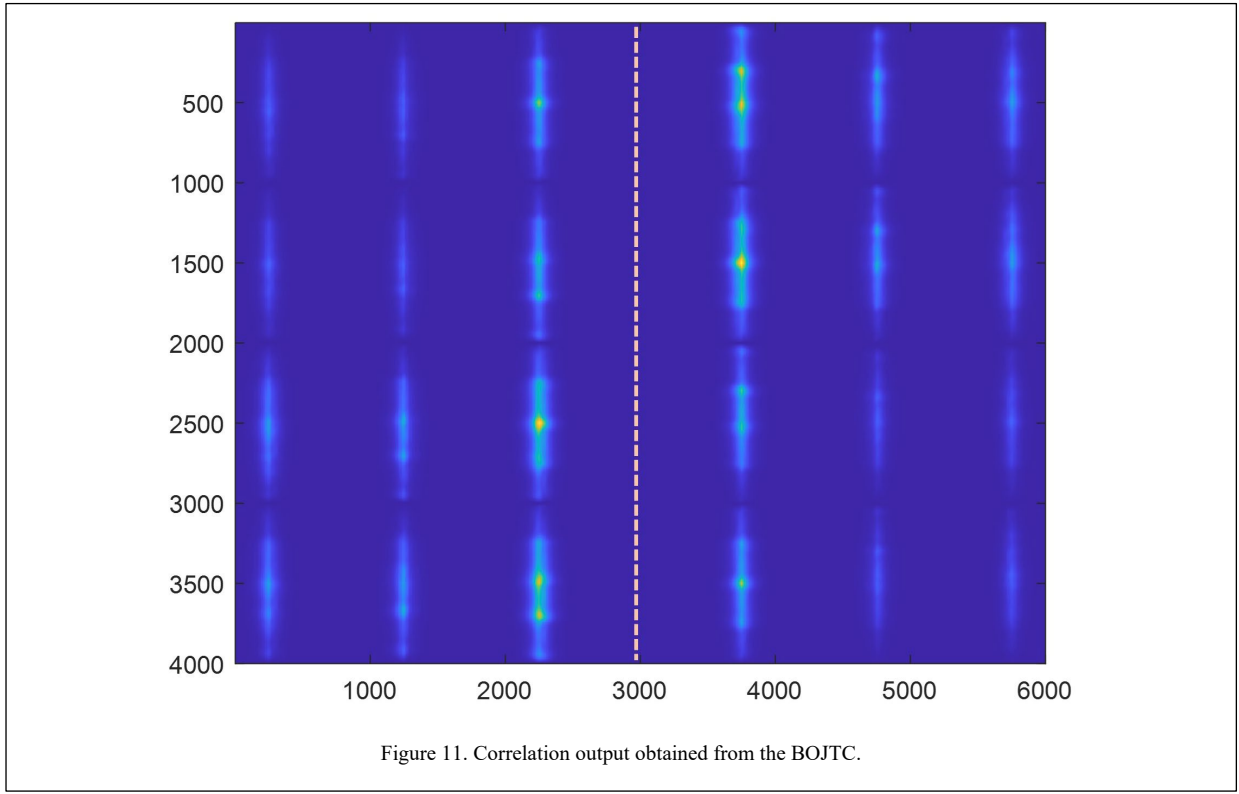

Figure 11. Correlation output obtained from the BOJTC.

## 5. Conclusion

This research has successfully demonstrated the theory and simulated performance of the SPMT approach for multi-object SSRI detection. The simulated integration of this preprocessing technique with a BOJTC shows promising results, effectively identifying objects regardless of their shift, scale, or rotation transformation.

As we conclude the theoretical and simulation phases of this research, our next steps will focus on validating these findings through experimental implementation. We plan to construct a physical prototype of the proposed system to evaluate its performance under real-world conditions. Additionally, an SOPP system can be constructed by replacing the original single lens in an OPP with a micro-lens array for parallel segmentation. This experimental setup will incorporate the micro-lens array for simultaneous multi-region FTs, along with the necessary optical and electronic components, and digital LPT calculation.

A primary objective of our future work is to explore the potential for ultra-fast multi-object parallel correlation using hybrid opto-electronic correlators. This setup offers significant advantages for applications requiring real-time processing of complex scenes with multiple objects involved. We anticipate that the parallel processing capabilities of these hybrid systems will substantially reduce the computational burden compared to traditional digital approaches.

Several aspects will be specifically investigated in the experimental phase: the feasibility of the system to handle larger numbers of objects and more complex scenes; the detection accuracy limits across the shift scale and rotation invariant detection; the maximum achievable processing speed of hybrid opto-electronic correlators. Through this experimental validation, we aim to establish the practical viability of our approach and quantify its performance advantages over existing methods.

**Acknowledgements**

The work reported here was supported by the Air Force Office of Scientific Research under Grant Agreements No. FA9550-18-01-0359 and FA9550-23-1-0617.